\documentclass[12pt]{iopart}
\usepackage{latexsym}

\begin{document}\jl{1}

\title[Unitary Dynamics]{Can We Believe in a
Purely Unitary\\ Quantum Dynamics?}

\author{F Herbut\footnote[1]{E-mail:
fedorh@infosky.net}}
\address{Serbian Academy of
Sciences and Arts, Knez Mihajlova 35, 11000
Belgrade, Serbia and Montenegro}

\date{\today}

\begin{abstract}
It is pointed out that the question of a purely
unitary quantum dynamics amounts to the question
if von Neumann entropy of a dynamically closed
quantum system is preserved in evolution.
\end{abstract}

\maketitle \normalsize \rm

Without Schlosshauer's recent critical analysis
of relevant experiments \cite{Schloss} showing
that the thesis from the title is not refuted
sofar the question from the title could not be
asked. The author of these lines proposes a "YES"
answer. This short note should give some
theoretical reasons for it.

Some remarkable results of Davies \cite{Davies}
(see Theorems 3.1 and 3.4 in Chapter 2 there) can
be viewed, as easily seen, in the following
manner:\\

{\it Purely unitary dynamics of a quantum system
that does not interact with its environment {\bf
is equivalent to} the following four assumptions:

(i) The dynamical law can be expressed as a
single-valued and time-dependent {\bf map} in the
set of density matrices of the quantum system.

(ii) The map {\bf preserves mixtures}
(mathematically: preserves convex combinations).

(iii) The map {\bf preserves pureness} of a
state.

(iv) The map is {\bf continuous in time}.}\\

Once this is recognized, the question from the
title can be changed as follows: {\it Is There
Theoretical Reason to Believe in the Four
Mentioned Requirements?}

If one believes that density operators and only
they express quantum states, then requirement (i)
sounds natural. An alternative to (i) would be
allowing two different mixtures making up the
same density matrix to evolve dynamically into
different density matrices. Gisin has shown
\cite{Gisin} that this brings us into conflict
with with the special relativity theory (which,
in turn, deserves believing in).

Density matrices may describe ensembles, and
these can be thought of as being suitable sets of
individual quantum systems. The latter evolve
each separately, and violation of requirement
(ii) would, as easily seen, make nonsense of this
idea (in view of the fact that sets can be
thought of as being composed of subsets in
different subjective ways).

Requirement (iii) can be understood as saying
that once we have complete information on a
system, without interaction, we will never lose
it.

Violation of requirement (iv) would introduce
discontinuities in the evolution, and these would
be hard to understand in view of the homogeneity
of the time axis.\\

It seems to me that the given discussion of
requirement (iii) needs elaboration.

A modified view of Davies' theorems would keep
requirements (i), (ii), and (iv) intact, and
replace (iii)  by the following:\\

{\it (iii)' The von Neumann {\bf entropy does not
change} in evolution of a dynamically isolated
quantum system.}\\

It is obvious that (iii)' implies (iii).
Conversely, (iii), in conjunction with the other
three requirements, gives unitary evolution due
to Davies' theorems, and unitary evolution
preserves entropy, i. e., (iii)' is valid.

Thus, to my mind, the question from the title
boils down to asking ourselves if we can believe
in preservation of entropy, unless interaction
changes it. The alternative is spontaneous change
of entropy. (Peres has pointed out \cite{Peres}
that entropy can even decrease in non-unitary
dynamics.)

I believe in the idea that if entropy changes,
one may look for the interaction that causes it.
Take the interesting article by Partovi
\cite{Partovi}, in which, following the important
observation in the preceding literature (see {\it
ibid.}) that entropy increase can only occur for
open systems, it is demonstrated that purely
unitary quantum dynamics does imply the second
law of thermodynamics. The way I see it, in
interaction of parts of a dynamically closed (or
isolated) system, the entropy of the parts may
increase, though that of the whole is unchanged,
on account of the quantum correlations created in
interaction. The correlations carry negentropy,
i. e., entropy with a negative sign in the
decomposition of the entropy of the whole. This
negentropy is called "mutual information" in
bipartite systems (see, e. g., \cite{FH} and the
references therein), and it has its
straightforward generalizations in multipartite
systems .

As it is well known, purely unitary quantum
dynamics implies the so-called measurement
paradox. This is a problem, and it must be
solved. To my mind, switching over to the
alternative (violation of the unitary law) is
seeking escape from the difficult search for
a solution.\\

\noindent {\bf References}\\

\end{document}